\documentclass[reprint,pra,aps,notitlepage,nofootinbib]{revtex4-2}
\usepackage{amsmath}
\usepackage{amssymb}
\usepackage{bm,braket}
\usepackage{float}

\usepackage[com pat=1.1.0]{tikz-feynman}
\usepackage{hyperref}
\usepackage{multirow}
\usepackage{subfig}
\hypersetup{
     colorlinks   = true,
     citecolor    = blue
}
\def\be{\begin{equation}}
\def\ee{\end{equation}}
\def\bea{\begin{eqnarray}}
\def\eea{\end{eqnarray}}
\def\bes{\begin{eqnarray}}
\def\ees{\end{eqnarray}}
\def\bi{\begin{itemize}}
	\def\ei{\end{itemize}} % ------- Define Greek Lowercase --------
	
\usepackage{amsthm}
\theoremstyle{definition}

\usepackage{tikz}
\usetikzlibrary{braids}
\begin{document}
\title{Non-Abelian Anyons with Rydberg Atoms}
	\author{Nora M.\ Bauer}
	\email{nbauer1@vols.utk.edu}
\affiliation{Department of Physics and Astronomy, The University of Tennessee, Knoxville, TN 37996-1200, USA}
	\author{Elias Kokkas}
	\email{ikokkas@vols.utk.edu}
\affiliation{Department of Physics and Astronomy, The University of Tennessee, Knoxville, TN 37996-1200, USA}
%\author{Aaron Bagheri}
%\email{bagheri@math.ucsb.edu}
%\affiliation{Department of Mathematics, University of California, Santa Barbara, CA 93106, USA}
\author{Victor Ale}
	\email{vale@vols.utk.edu}
\affiliation{Department of Physics and Astronomy, The University of Tennessee, Knoxville, TN 37996-1200, USA}
\author{George Siopsis}
	\email{siopsis@tennessee.edu}
\affiliation{Department of Physics and Astronomy, The University of Tennessee, Knoxville, TN 37996-1200, USA}
%\author{Zhenghan Wang}
%\email{zhenghwa@microsoft.com}
%\affiliation{Microsoft Station Q and Department of Mathematics, University of California, Santa Barbara, CA 93106, USA}
\date{\today}
\begin{abstract}
We study the emergence of topological matter in two-dimensional systems of neutral Rydberg
atoms in Ruby lattices. While Abelian anyons have been predicted in such systems, non-Abelian
anyons, which would form a substrate for fault-tolerant quantum computing, have not been generated. To generate anyons with non-Abelian braiding statistics, we consider systems with mixed-boundary punctures. We obtain the topologically distinct ground states of the system numerically using the iDMRG technique. We discuss how these topological states can be created using ancilla atoms of a different type. We show that a system with $2N+2$ punctures and an equal number of ancilla atoms leads to $N$ logical qubits whose Hilbert space is determined by a set of stabilizing conditions on the ancilla atoms. Quantum gates can be implemented using a set of gates acting on the ancilla atoms that commute with the stabilizers and realize the braiding group of non-Abelian Ising anyons. 

\end{abstract}
\maketitle

\section{Introduction}\label{sec:I}
Systems in two spatial dimensions exhibiting topological order \cite{bib:1}  are promising candidates for fault-tolerant quantum computers and quantum memory \cite{bib:2,bib:3}. This exotic state of matter supports quasi-particle excitations which obeys anyonic quantum statistics. Anyons come in two types: Abelian and non-Abelian. Abelian anyons cannot be used for quantum computation since their braiding produces only a phase \cite{bib:4}. Nevertheless, they can be used to store information in a topologically protected way. Many quantum error correcting schemes involve Abelian anyons. In Kitaev's toric code \cite{bib:3} three Abelian anyonic excitations emerge by placing a square lattice spin system on a torus. It was later realized that toroidal geometry was not a necessary requirement, which led to generalizations such as the surface or planar code \cite{bib:5,bib:6}. On the other hand, braiding non-Abelian anyons changes their state by a unitary matrix.  For both types of anyons error protection against local perturbations is guaranteed by the robust ground state degeneracy and the energy gap between ground state and excited states \cite{bib:7}.

Topological phases of matter have been under enormous investigation because of their importance in condensate matter physics as well as their potential applications to the field of quantum information and quantum computation. Despite that, the experimental realization of topological ordered states  has been elusive. Abelian anyons have been experimentally  observed in fractional quantum Hall effect (FQHE)  at filling $\nu=\frac{1}{3}$ \cite{bib:8}, whereas the existence  of non-Abelian anyons at fillings $\nu=\frac{5}{2}$ and $\nu=\frac{12}{5}$ has yet to be confirmed. Alternative approaches involve spin systems in various lattices.  It was theoretically predicted in Ref.\ \cite{bib:9} that a spin system in a Honeycomb lattice supports Abelian and non-Abelian anyons, the latter only in the presence of a magnetic field.   Recently, neutral Rydberg atoms placed on Kagome and Ruby lattices were proposed as candidates for quantum computing and quantum memory \cite{bib:10,bib:11}.
Numerical and experimental results \cite{bib:12,bib:13} claim the realization of a topological spin liquid using neutral atoms and the mechanism of the Rydberg blockade \cite{bib:14,bib:15}. More specifically, previous results suggest the emergence of the $\mathbb{Z}_2$ topological order of the toric code. This is an Abelian anyon model described by the vacuum $\mathbb{I}$ and the excitations  $e$, $m$ and $\epsilon$, which can be engineered to create topologically protected quantum memory. $e$ and $m$ are bosons, whereas their composition $\epsilon =e \times m$ is a fermion.
%The emergence of $e$ and $m$ abelian anyons was demonstrated by measuring the Wilson and 't Hooft string operators.

To achieve fault-tolerant quantum computation, we need non-Abelian braiding statistics.  Attempts to obtain non-Abelian statistics from systems with Abelian statistics using topological defects have been studied extensively. Such defects  can be twists \cite{bib:16,bib:17,bib:18,bib:19} or punctures \cite{bib:20,bib:21} on the lattice.

In this work, we consider Rydberg atoms on the Ruby lattice, which was studied in \cite{bib:12,bib:13}, and introduce punctures with mixed boundaries.  There are two types of boundaries depending on the  type of anyon condensate. For $e$ anyons we have a smooth boundary, whereas for $m$ anyons we have a rough boundary. We obtain the ground state of the system numerically using the iDMRG method  developed in \cite{bib:22}. We discuss how these topological states can be created and controlled with the aid of ancilla
atoms of a different type than the ones forming the Ruby lattice. We show that a system with $2N + 2$ mixed-boundary punctures, and an equal number
of ancilla atoms for control, leads to $N$ logical qubits. Their Hilbert space is obtained by constraining the state of the ancilla atoms using a set of stabilizing
conditions. Quantum gates can be implemented using a set of gates acting on
the ancilla atoms. The latter commute with the stabilizers and form a realization of the braiding group of non-Abelian Ising anyons.

Our discussion is organized as follows. In  Section \ref{sec:II}, we introduce the model we use for the Rydberg atoms and analyze the  ground states for a system with two mixed-boundary punctures.  Section \ref{sec:III} focuses on numerical calculations using  iDMRG,  where we first confirm that we obtain the spin liquid phase, and then show results for the ground states of the system.   In Section \ref{sec:IV}, we introduce a logical qubit basis on a system containing mixed-boundary punctures with the aid of stabilizer contraints on the ancilla atoms. We discuss how interactions with ancilla atoms can be used to prepare states and apply quantum gates on the system. We show that quantum gates can be implemented with the aid of a set of gates acting on
the ancilla atoms that commute with the stabilizers and realize the braiding group of non-Abelian
Ising anyons.  Finally in Section \ref{sec:V}, we summarize our results. 

\section{The Model}\label{sec:II}
%We studied the emergence of non-abelian anyons in a system with Rydberg atoms by creating punctures with mixed boundaries on the lattice. 

We consider ${}^{87} \rm{Rb}$ neutral atoms  placed on the sites of a Ruby optical lattice of lattice spacing $a$ governed by the Hamiltonian 
\begin{equation} \begin{split}
    H &= \frac{\Omega(t)}{2} \sum_{i} (e^{-i \alpha(t) } b_{i} + e^{i \alpha(t)} b_{i}^\dagger) - \sum_{i} \Delta_{i}(t) n_{i} \\
    &+ \frac{1}{2} \sum_{i,j} V(r_{ij}) n_i n_j  \ ,
\end{split}
\label{eq:1}
\end{equation} 
with each site containing at most one atom. Equivalently,  we can consider the atoms located on the links of a Kagome lattice, since these sites make up the vertices of the Ruby lattice. The system is driven by coherent laser beams. The atom at the site $i$ can be either in the ground state $\ket{g}_i$ or the Rydberg state  $\ket{r}_i$, with $\Omega (t)$ denoting the Rabi frequency of oscillation between the two states, which is a global parameter controlled by varying the laser intensities \cite{bib:13}.  Excitations from the ground state to the Rydberg state are driven by the laser detuning $\Delta_i (t)$ which can be adjusted on individual atoms, as needed. The operators $b_i$ and $b_i^\dagger$ are bosonic lowering and raising operators for the two-level quantum system $\{ \ket{g}_i,  \ket{r}_i \}$ at site $i$ ($b_i = \ket{g}_i \bra{r}$). The particle number operator $n_i= b_i^\dagger b_i = \ket{r}_i\bra{r}$, determines if the atom at site $i$ is excited by projecting that atom onto the Rydberg state. For each site $i$, we can construct the local Pauli operators $X_i= b_i + b_i^\dagger = \ket{g}_i\bra{r}+\ket{r}_i\bra{g}$, $Y_i= i(b_i - b_i^\dagger) = i\ket{g}_i\bra{r}-i\ket{r}_i\bra{g}$, and $Z_i= b_ib_i^\dagger - b_i^\dagger b_i = \ket{g}_i\bra{g}-\ket{r}_i\bra{r}$. The latter can also be written in terms of the number operator $n_i$ as $Z_i= \mathbb{I} -2n_i$. $\alpha (t)$ is a global phase factor. The case $\alpha=0 $ is directly accessible with the corresponding term in the Hamiltonian being the Pauli matrix $ X$. To introduce a non-vanishing $\alpha $, we apply time evolution with a Hamiltonian in which the detuning is dominant ($H\approx - \Delta \sum_i n_i $), and adjust the global detuning $\Delta$ to obtain the desired phase $\alpha$, as can be seen using  $e^{-i \alpha Z } X e^{i \alpha Z } = \cos \alpha X + i \sin \alpha  Y$. The phase $\alpha $ can be varied in time by adjusting the detuning.    Finally, there is a strong  repulsion due  to  the van der Waals potential $V(r_{ij})  = \Omega ({R_b}/{r_{ij}})^6 $, where $R_b$ is the Rydberg blockade radius. Hence, for every atom in the Rydberg state, there is a barrier that blocks atoms inside a radius $R_b$ from getting excited. For our work, we will fine tune the lattice spacing $a$ and Rabi frequency $\Omega$ so that for each excited atom, its six closest neighbors are inside the blockade radius. 

Due to the Rydberg blockade, operators within a triangle in the Ruby lattice are not independent of each other. Consider a triangle formed by the set of sites $\mathcal{T} = \{ i_1,i_2,i_3 \}$. There are four possible states, the three Rydberg states $\ket{r}_j$ ($j\in\mathcal{T}$), and the ground state $\ket{g}_{\mathcal{T}}$ with all atoms in their respective ground states. In the basis $\{ \ket{g}_{\mathcal{T}}, \ket{r}_{i_1} , \ket{r}_{i_2} , \ket{r}_{i_3}  \}$, operators can be thought of as $4\times4$ matrices. For example, the part of the Hamiltonian proportional to the Rabi frequency restricted to the triangle in the Ruby lattice under consideration can be written as
\be\label{eq:HT} H_{\mathcal{T}} = \frac{\Omega}{2} \sum_{i\in\mathcal{T}} (e^{-i \alpha } b_{i} + e^{i \alpha} b_{i}^\dagger) = \frac{\Omega }{2} \left( \begin{array}{cccc}
         0& e^{i\alpha} &e^{i\alpha} &e^{i\alpha}  \\
         e^{-i\alpha} & 0 & 0 &0 \\
         e^{-i\alpha} & 0&0 &0 \\
         e^{-i\alpha} &0 &0 &0
    \end{array} \right) \ee 
This result will be useful in the calculation of non-local string operators to which we turn next.

 To study the phase of the system, we consider the non-local topological string operators $\bm{X}_S =\prod_{i \in S } X_i$ and $\bm{Z}_{S'} =\prod_{i \in S' } Z_i$, along the strings $S$ and $S'$, respectively. These operators were introduced in \cite{bib:9} for an exactly solvable $\mathbb{Z}_2$ lattice gauge theory.  In the context of gauge theories, the $\bm{X}_S$ and $\bm{Z}_{S'}$ string operators are the Wilson and 't Hooft lines, respectively. Here, they will be used to explore the different phases of matter that our system exhibits.

\begin{figure}[t]
    \centering
\includegraphics[width=0.45\textwidth]{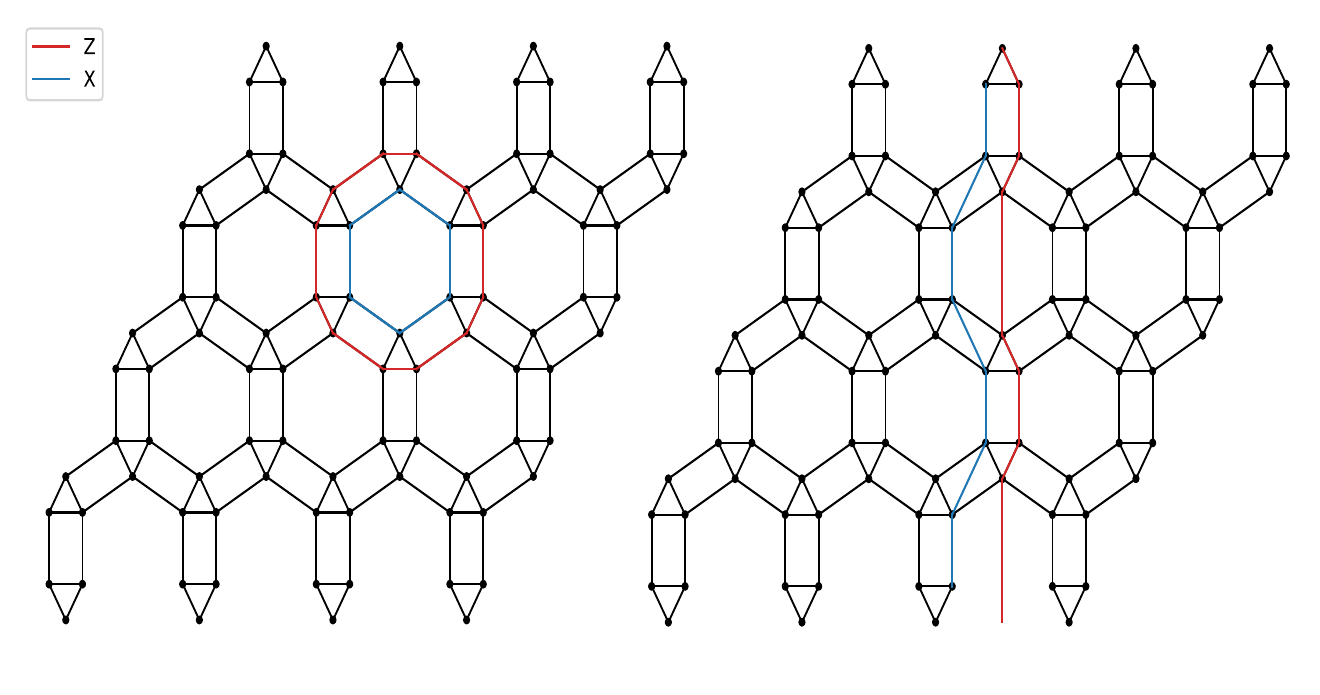}
    \caption{Left: Ruby lattice with open boundary conditions. Closed $\bm{Z}$ and $\bm{X}$ loop string operators are used to detect the phase of the system. Right: Ruby lattice with periodic boundary conditions (cylinder). Closed $\bm{Z}$ and $\bm{X}$ string operators circling the cylinder are used to detect the phase of the system. For the periodic lattice, loop string operators can also be used.} 
    \label{fig:II1}
\end{figure}

As observed in \cite{bib:12,bib:13}, this Rydberg atom model has a trivial phase for small $\Delta / \Omega$, a valence bond solid (VBS) phase for large $\Delta / \Omega$, and a quantum spin liquid (QSL) phase for intermediate values of $\Delta / \Omega$. We identify the phase of our system by measuring closed loops for both $\bm{X}$ and $\bm{Z}$ topological string operators, as shown in Figure \ref{fig:II1}. 
%S. The $Z$ operator is defined as the parity operator measured along the sites in S: $Z = \prod_{i \in S} \sigma_i^z $. The X operator is measured by first implementing a unitary transformation \cite{6} by time evolving the system with Hamiltonian $H'$ 
%\begin{equation} 
%H' = \frac{\Omega '}{2} \sum_\textbf{i} (b_\textbf{i} + b_\textbf{i}^\dagger) + \frac{1}{2} \sum_\textbf{i,j} V'(r_{\textbf{i}, \textbf{j}}) n_\textbf{i} n_\textbf{j} 
%\end{equation} 
%for time period 
%\begin{equation} 
%\tau = \frac{4}{3 \sqrt{3} \Omega } 
%\end{equation} 
%and then measuring the Z operator along the sites dual to the off-diagonal X string sites. 
Specifically, a closed $\bm{Z}$ loop is vanishing in the trivial phase and non-zero in the QSL and VBS phases, whereas a closed $\bm{X}$ loop is finite in the trivial and QSL phases and vanishes in the VBS phase of matter. The QSL phase realizes Kitaev's toric code  $\mathbb{Z}_2$ topological order, as both $e$ and $m$ anyons emerge. The numerical results are discussed in more detail in Section \ref{sec:III}.

On the other hand, open string operators are associated with the  creation of these Abelian anyons. A pair of $m$ anyons is created at the endpoints of an open $\bm{Z}$ string, whereas a pair of $e$ anyons is created at the endpoints of an open $\bm{X}$ string operator.

The boundaries of the Ruby lattice in Figure \ref{fig:II1} can be either periodic or open. The former choice gives the toric code \cite{bib:3}, which has four-fold ground state degeneracy. The latter gives the planar or surface code \cite{bib:5,bib:6}, which is experimentally more feasible. In the planar code, the boundaries can be either rough or smooth. The type of the boundary, in the planar code, defines the type of anyon condensation. Smooth (rough) boundaries contain $e$ ($m$) anyon condensation.

\begin{figure}[ht!]
    \centering
\includegraphics[width=0.45\textwidth]{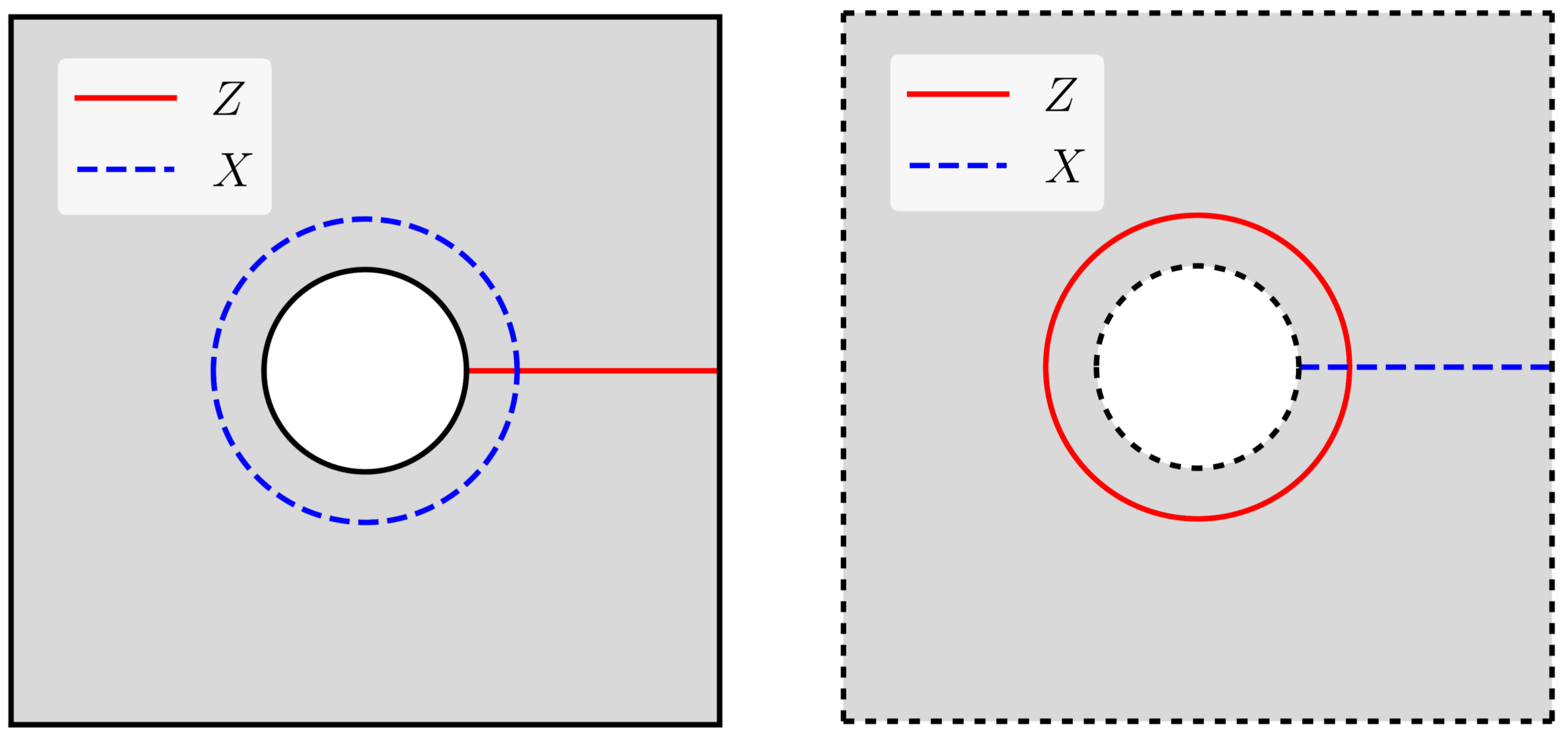} 
    \caption{Left: Planar code with rough boundaries ($m$-condensation), gives the $\ket{\mathbb{I}}$ and $\ket{m}$ states.
    Right: Planar code with smooth boundaries ($e$-condensation), gives the $\ket{\mathbb{I}}$ and $\ket{e}$ states.} 
    \label{fig:II2}
\end{figure} 

The dimensionality of the  planar code's Hilbert space $\mathcal{H}$ depends on the boundaries and the genus of the plane. For a genus-0 plane, we need alternating boundaries in order to encode a logical qubit $(dim(\mathcal{H})=2)$. For a genus-1 plane, we can encode a qubit using uniform boundaries, as shown in Figure \ref{fig:II2}. Uniform rough boundaries, which correspond to an $m$ condensate, give 2 ground states denoted by $\ket{\mathbb{I}}$ and $\ket{m}$. Similarly, smooth boundaries, which correspond to an $e$ condensate, give 2 ground states denoted by $\ket{\mathbb{I}}$ and $\ket{e}$.  Such systems, supporting $e$ and $m$ Abelian anyons, can be used for quantum memory or non-universal quantum computation. To construct a four-dimensional Hilbert space spanned by states $(\ket{\mathbb{I}}, \ket{e}, \ket{m}, \ket{\epsilon})$, where $\ket{\epsilon}$ is obtained by fusing $e$ and $m$ anyons, we need a puncture and mixed boundaries in both lattice and puncture. 

\begin{table}[b!]
\begin{ruledtabular}
\begin{tabular}{lcccc}
\textrm{State}&
$\bm{Z}_C$&
$\bm{X}_{C'}$&
$\bm{Z}_{S}$&
$\bm{X}_{S'}$\\
\colrule
$\ket{\mathbb{I}}$ & 1 & 1 & 0 & 0 \\
$\ket{e}$ & -1 & 1 & 0 & 0 \\
$\ket{m}$ & 1 & -1 & 0 & 0 \\
$\ket{\epsilon}$ & -1 & -1 & 0 & 0 \\
$\ket{+}$ & 0 & 0 & 1 & 1 \\
$\ket{-}$ & 0 & 0 & -1 & -1
\end{tabular}
\end{ruledtabular}
\caption{Ideal expectation values of four loop operators, {$\bm{Z}_C$}, $\bm{X}_{C'}$, $\bm{Z}_{S}$, {$\bm{X}_{S'}$}, defined in Figure \ref{fig:II3}, to detect the four ground states $ \ket{\mathbb{I}}, \ket{e}, \ket{m}, \ket{\epsilon} $, and the superposition states $\ket{\pm}$ (Eq.\ \eqref{eq:2a}) of a lattice with two mixed-boundary punctures. \label{tab:1}%  
}
\end{table} 

\begin{figure}[t!]
    \centering 
\includegraphics[width=0.45\textwidth]{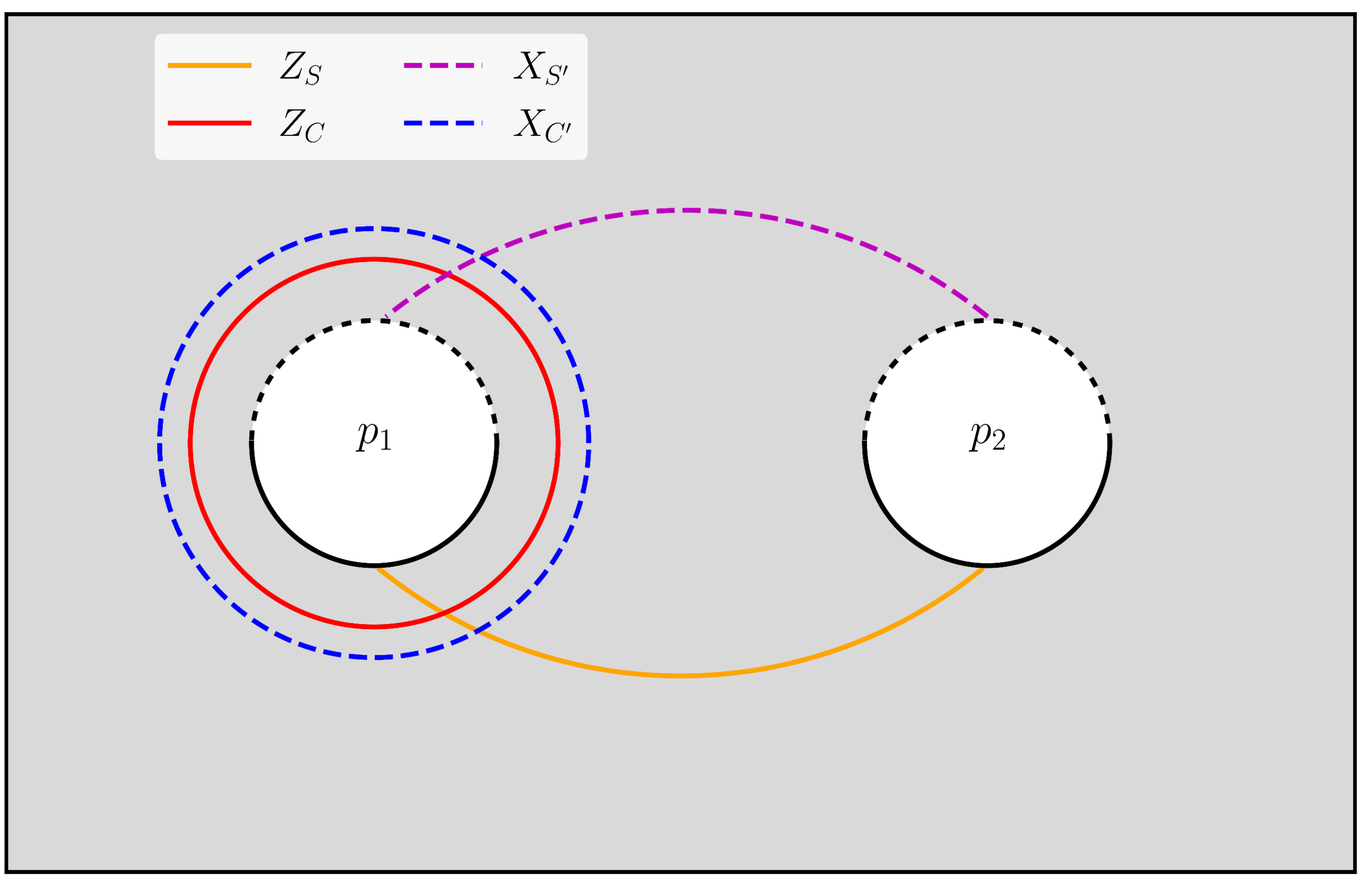}  
    \caption{Two mixed-boundary punctures on a lattice with rough boundary conditions. The loops enclosing a puncture and the strings connecting the two punctures define operators that are used to detect the degenerate ground states. } 
    \label{fig:II3} 
\end{figure}

Since the mixed boundary punctures can condense  both $e$ and $m$ anyons on their boundaries, we can  construct  the set of  ground states  $\{ \ket{\mathbb{I}}, \ket{e}, \ket{m}, \ket{\epsilon} \} $ using two mixed-boundary punctures, $p_1$ and $p_2$. We define four non-local string  operators that act on this system, {$\bm{Z}_{C}$}, {$\bm{X}_{C'}$}, {$\bm{Z}_{S}$}, and {$\bm{X}_{S'}$}, as shown in Figure \ref{fig:II3}. $C$ and $C'$ are loops around puncture $p_1$ (or equivalently $p_2$). $S$ and $S'$ are strings connecting the rough ($m$) and smooth ($e$) boundaries of the two punctures, respectively. We can use these four operators to detect the ground state. The loop operators $\bm{Z}_C$ and $\bm{X}_{C'}$ are independent of the choice of loops $c$ and $C'$, as long as they enclose a puncture.  The ideal values of these operators for the four ground states, as well as the superposition states $\ket{\pm}=(\ket{e}\pm \ket{m})/\sqrt{2 } $ (Eq.\ \eqref{eq:2a}), are given in Table \ref{tab:1}.

% Since the mixed boundary punctures can condense  $e$ and $m$ anyons on the boundaries, a system with 2 mixed boundary punctures could support the following 2 states of interest: (1) $\ket{e}$  states where the $e$-condensed boundaries are connected by a $Z$-string and enclose an $e $ anyon (2) $\ket{m } $  states where the $m $-condensed boundaries are connected by an $X $-string and enclose an $m $ anyon . 

%We now consider preparing the $\ket{e }_{12 }  $ and $\ket{m }_{12 }  $ ground states in the Ruby lattice model with punctures.
To better understand mixed-boundary punctures, it is instructive to consider the case of two punctures with uniform ($e $ or $m $ condensed) boundaries. For $e $ condensed boundaries, $e $ anyons can be present within the 2 punctures, and their presence is detected by measuring the $\bm{Z}_C $ operator along a loop $C$ containing one of the puncture. If  $\braket{\bm{Z}_C }=-1 $, an $e $ anyon is enclosed in the puncture, whereas $\braket{\bm{Z}_C }=1 $ implies that no anyon is enclosed. For a system with two punctures with $m$ condensed boundaries, the presence of an $m $ anyon within a puncture is determined by the sign of the $\bm{X}_{C'} $ operator along a loop $C'$ surrounding the puncture, with $\braket{\bm{X}_{C'} }=-1,1 $ implying that an $m $ or no anyon is enclosed in the puncture, respectively. 
 
Additionally, in a system with two punctures of $e $ condensed boundaries, we consider the puncture-to-puncture operator  {$\bm{X}_{S}$}, along a  string $S$ connecting the boundaries of the two punctures. $S$ is an open string with $e$ anyons at its two ends. The expectation value of this string is the overlap of the ground state with the state resulting from adding an $e $ anyon to each puncture. An analogous operator  {$\bm{Z}_{S'}$} along a string $S'$ connecting the $m$ boundaries of the two punctures is measured in a system with $m$ condensed boundaries. 
%and $\braket{\bm{X}^L_C }=1 $ meaning no anyon is enclosed.

\begin{figure}[ht!]
\centering
\subfloat[Two smooth boundary punctures on a cylinder.]{
	\label{subfig:notwhitelight}
	\includegraphics[width=0.45\textwidth]{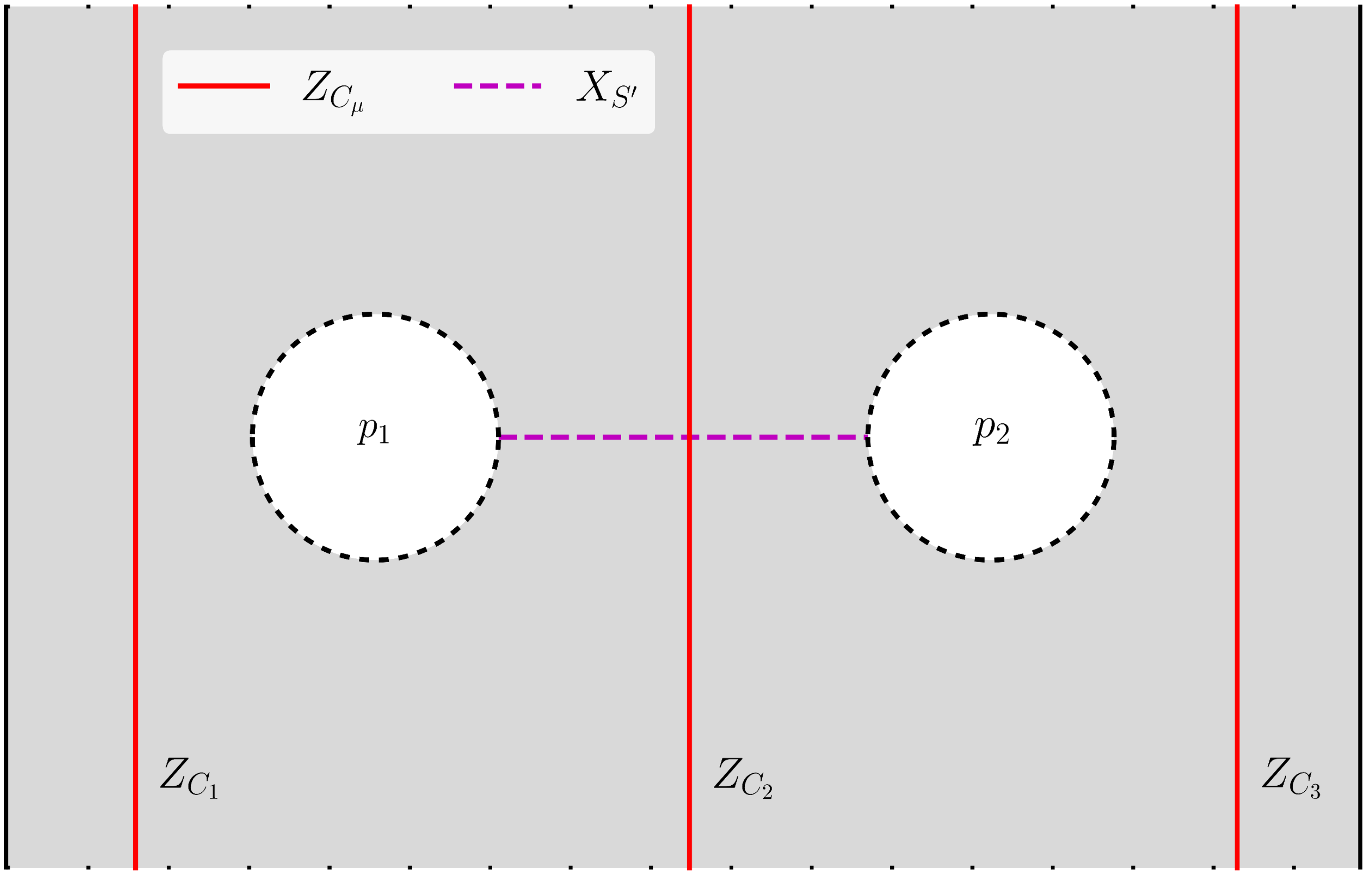} } 
	
 \subfloat[Two rough boundary punctures on a cylinder. ]{
	\label{subfig:correct}
	\includegraphics[width=0.45\textwidth]{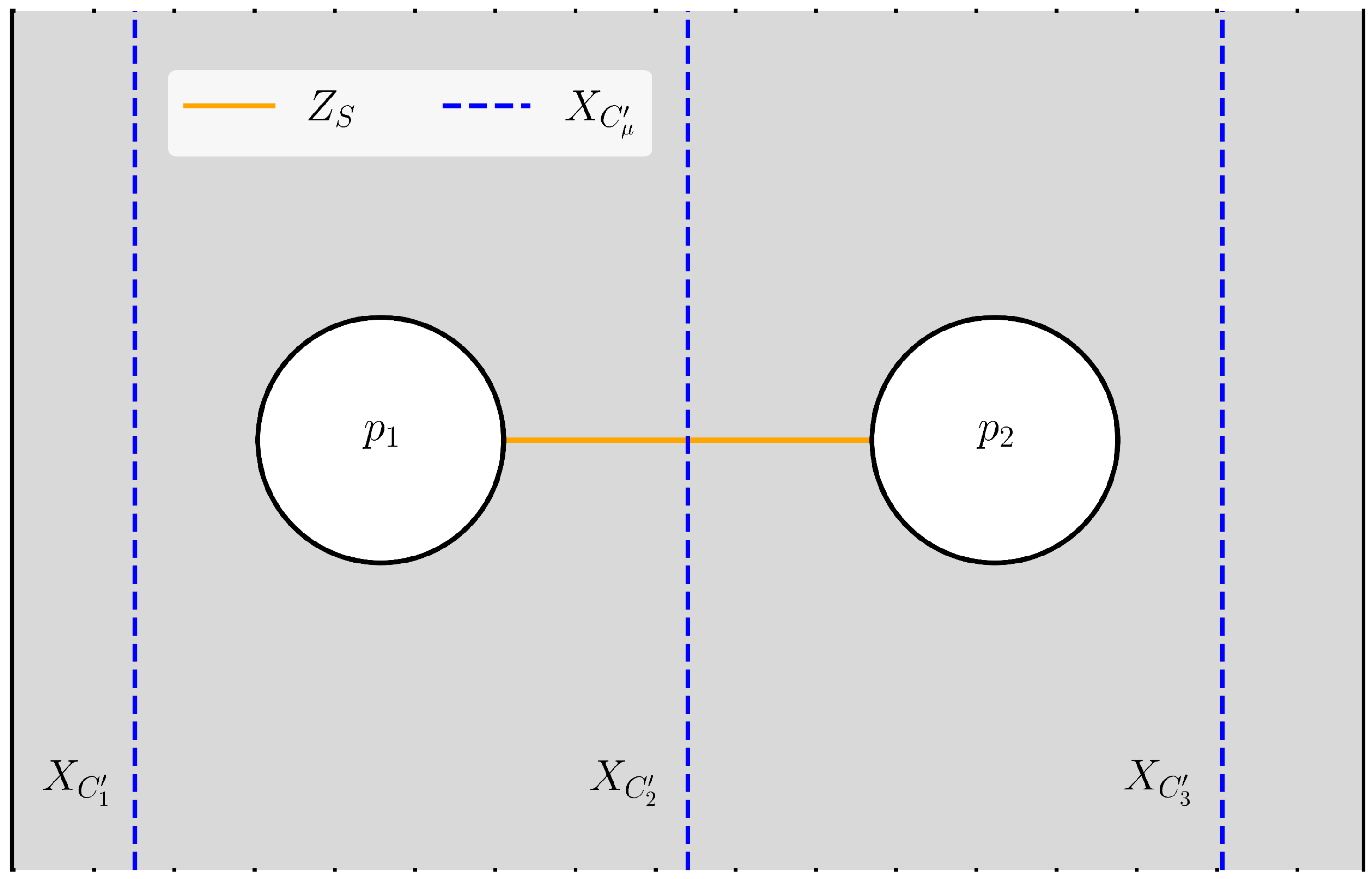} } 

\caption{In (a), an $\bm{X}$ loop around each puncture is effectively given by the product of two vertical strings across the puncture. If an $m$ anyon is present, the product is negative, otherwise it is positive. In (b), a $\bm{Z}$ loop around the puncture is given by the product of two vertical strings. }
\label{fig:II4}

\end{figure}

For the numerical simulations discussed in Section \ref{sec:III}, we will also have to consider a lattice with periodic boundary conditions on the top and bottom sides of the lattice making the $y$ axis periodic. Thus, our system is defined on an infinite cylinder along the $x$ axis. To measure {$\bm{Z}_C$} and {$\bm{X}_{C'}$}, we take advantage of the periodic boundary conditions in the $y$-direction such that, rather than measuring a large string that encloses the entire puncture, we measure two strings around the circumference of the cylinder: one before and one after the puncture. A change of sign of the expectation value of {$\bm{Z}$} ({$\bm{X}$}) resulting from a measurement of the two strings before and after the puncture indicates that an $m$ ($e$) anyon is enclosed in the puncture. Therefore, we report on numerical results for {$\bm{Z}_C$} around the two punctures in terms of {$\bm{Z}_{C_1}$}, {$\bm{Z}_{C_2}$}, and  {$\bm{Z }_{C_3}$} with loops $C_1, C_2, C_3$ in the $y$-direction around the cylinder, as shown in Figure \ref{subfig:notwhitelight}. An identical construction is used for  {$\bm{X}_C$}, as shown in Figure \ref{subfig:correct}.  
%We measure loop operators around a puncture by considering loops wrapping around the cylinder in the periodic $y $ direction. Thus, if, for example, an $e$ anyon is enclosed, the sign of the $\bm{Z} $ loop should be opposite between loops drawn around the cylinder before and after the puncture in $x$ direction. These loops on a 2-puncture system with $e $ and $m $ boundary conditions are shown in Figure \ref{fig:II4}. 

\begin{figure}[b!]
    \centering 
\includegraphics[width=0.45\textwidth]{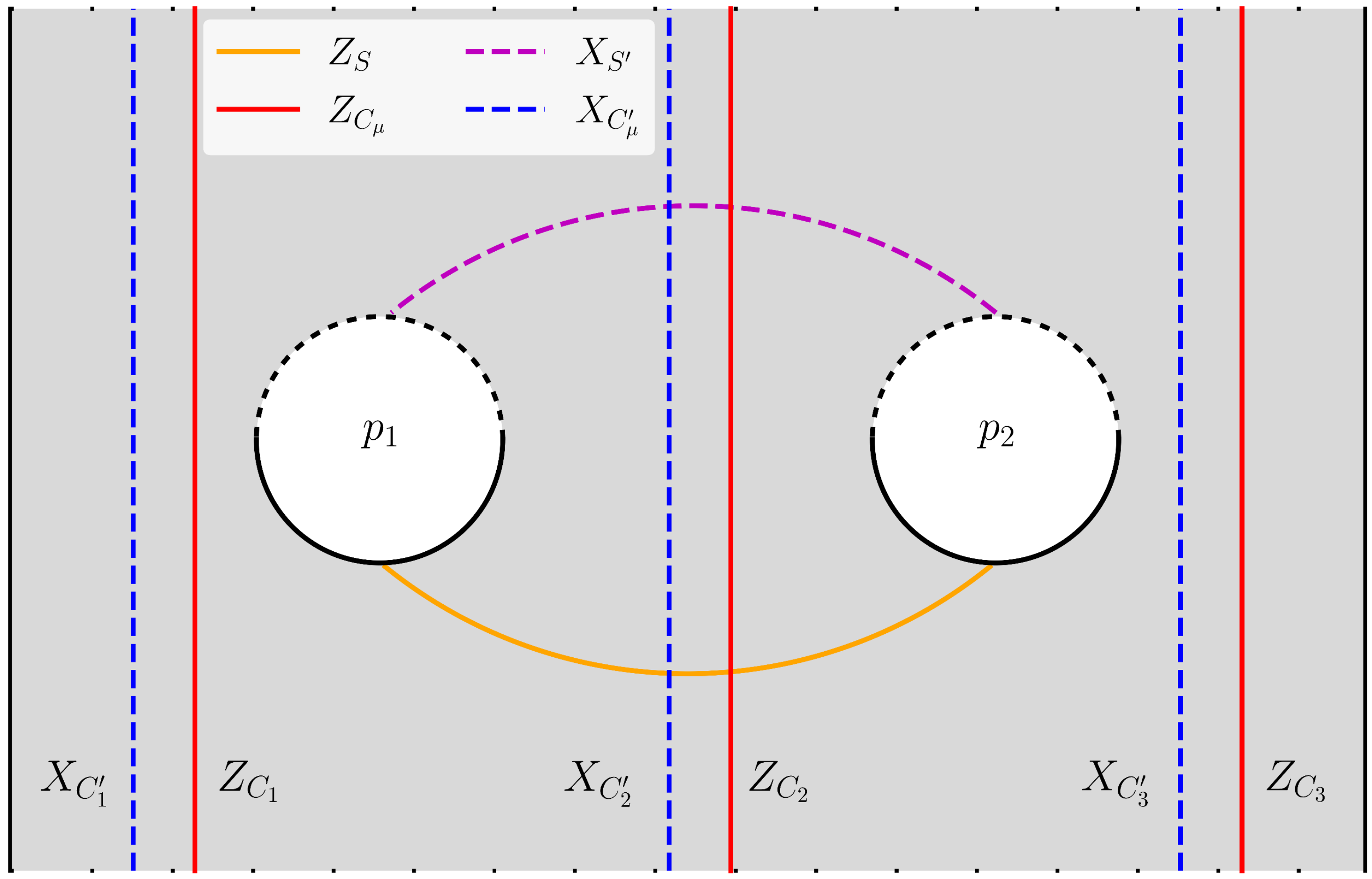} 
    \caption{Two mixed-boundary punctures on a lattice with periodic top and bottom boundary conditions. }
    \label{fig:II5} 
\end{figure}

Turning to the cylindrical system containing two mixed-boundary punctures, once again we can detect the four degenerate ground states, displayed on Table \ref{tab:1}, by measuring the four loop operators shown in Figure \ref{fig:II5}.

Following Ref.\ \cite{bib:6}, we introduce mixed boundary punctures in the lattice,  and seek to obtain non-Abelian statistics from Rydberg atoms.
To this end, an important ingredient is the creation of the superposition states 
\begin{equation} 
\ket{\pm }_{12}  = \frac{\ket{e }_{12 }  \pm \ket{m }_{12 }  }{\sqrt{2 } }   \ ,
\label{eq:2a}
\end{equation} 
for the two punctures $p_1$ and $p_2$, where the states $\ket{e }_{12 }  $  and $\ket{m }_{12 }  $ have $e$ and $m$ anyons, respectively, enclosed in the punctures. We obtained these states by running the iDMRG algorithm over many random initial states and then classifying them using the loop operators. 
% Now we prepare a system with 4 mixed boundary punctures, analogous to the system in \cite{bib:21} on the toric code. We can compute the expectation values of $Z $ and $X $ loops starting and ending on $Z $ and $X $ boundaries to confirm the mixed boundaries. 

\begin{figure}[ht!]
    \centering  
\includegraphics[width=0.45\textwidth]{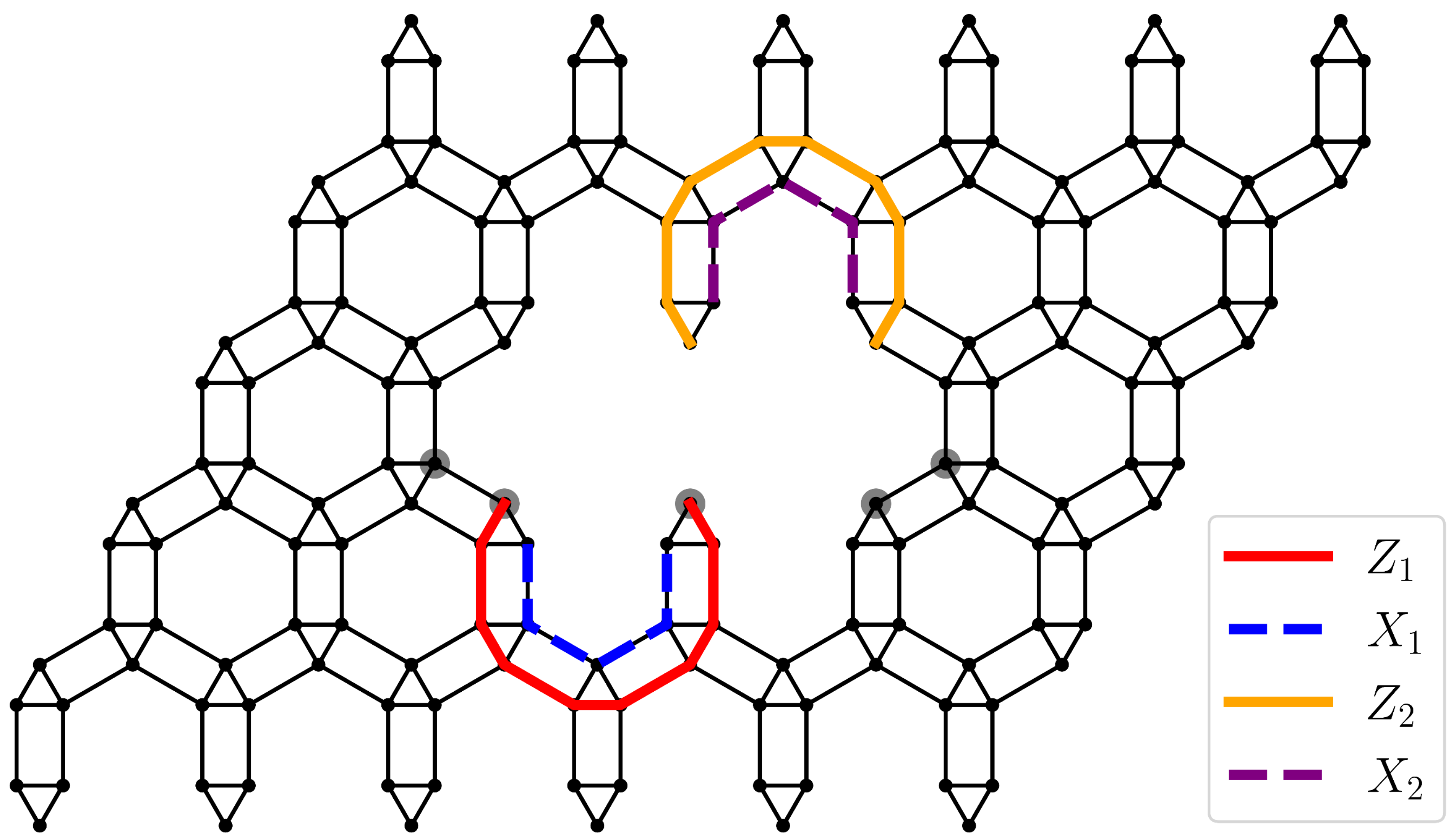} 
    \caption{Mixed-boundary puncture on the Ruby lattice. Sites highlighted in gray have detuning $\Delta' =0.48 \Delta$. The string operators terminating on the boundaries of the puncture are used to detect the condensation type. Expectation values are given in Table \ref{tab:2a}.} 
    \label{fig:IIlast} 
\end{figure}

The punctures were implemented by removing unit cells of atoms on a Ruby lattice. In the spin liquid regime, the boundaries naturally condense $m$ anyons, where a string loop starting and ending on the boundary has a finite $\langle \bm{Z} \rangle $ value and a vanishing $\langle \bm{X} \rangle $ value. Boundaries with $e$ anyon condensation can be obtained by decreasing the detuning to $\Delta '  < \Delta$ (as discussed in Section \ref{sec:III}, for numerical results, we set $\Delta' = 0.48 \Delta $) on the sites that make up the boundary of the punctures. Measurements of the string operators give vanishing $\langle \bm{Z} \rangle $ and non-vanishing $\langle \bm{X} \rangle $. The mixed-boundary punctures were created by only changing the detuning of the boundary sites on half of the boundary. Specifically, the detuning was reduced to $\Delta' =0.48 \Delta $ on highlighted sites shown in Figure \ref{fig:IIlast}.  
 % The expectation values $\langle \bm{Z} \rangle $ and $\langle \bm{X} \rangle $ for strings starting and ending on the two types of the mixed-boundary puncture are given in Table \ref{tab:2a}.  
  
% \begin{figure}[ht!]
%     \centering
% \includegraphics[width=0.4\textwidth]{loops.png%} 
%     \caption{Ruby lattice of Rydberg atoms with %four punctures. Hexagons with detuning $\delta$ %($\delta ' $) are highlighted in blue (red).}
%     \label{fig:1}
% \end{figure}

\section{Numerical Method and Results}\label{sec:III}

We obtained the ground state of the system with Hamiltonian \eqref{eq:1} on a Ruby lattice geometry numerically using iDMRG with the TenPy python package introduced in Ref.\ \cite{bib:11}. Following \cite{bib:13}, the parameters we used  were $\Omega = 2 \pi \times 1.4\times 10^{6}$, $\alpha=0$, $R_b = 2.4 a$ and  $R_{\text{trunc}} = \sqrt{7} a $,   with lattice spacing $a$ \cite{bib:7}. Here, $R_{\text{trunc}}$ is the truncation distance, which is the maximum distance between two sites for which the van der Waals interaction is included numerically in the Hamiltonian. Although the van der Waals potential is fixed in experiment, the Rydberg blockade radius $R_b$ can be effectively tuned by adjusting the Rabi frequency $\Omega$ and laser detuning $\Delta$ simultaneously  to alter the relative contribution of the van der Waals term  to the Hamiltonian. Then we performed iDMRG for different values of detuning $\Delta$, in order to obtain the phase diagram.

Expectation values for $\bm{Z}$ string operators were calculated directly using the ground states from iDMRG. Calculating the $\bm{X}$ string operators was not straightforward. Following Ref.\ \cite{bib:12}, starting from the ground state $\ket{\psi_0}$, we applied the time evolution operator $U_\tau = e^{-i\tau H'}$ under the Hamiltonian \eqref{eq:1} with $\Delta=0$ and $\alpha=-\pi/2 $,
\begin{equation} 
H' = \frac{i\Omega }{2} \sum_i (b_i - b_i^\dagger) + \frac{1}{2} \sum_{i,j} V(r_{ij}) n_i n_j   \ .
\label{eq:4}
\end{equation} 
for $\tau=\frac{4\pi}{3\sqrt{3} \Omega } $.
Then, we measured  the dual $\bm{Z}_S$ string operator on $\ket{\psi_\tau}= U_\tau \ket{\psi_0} $, which is equivalent to measuring an $\bm{X}_{S'}$ string on $\ket{\psi_0}$. 
\begin{figure}[ht!]
    \centering
\includegraphics[width=0.475\textwidth]{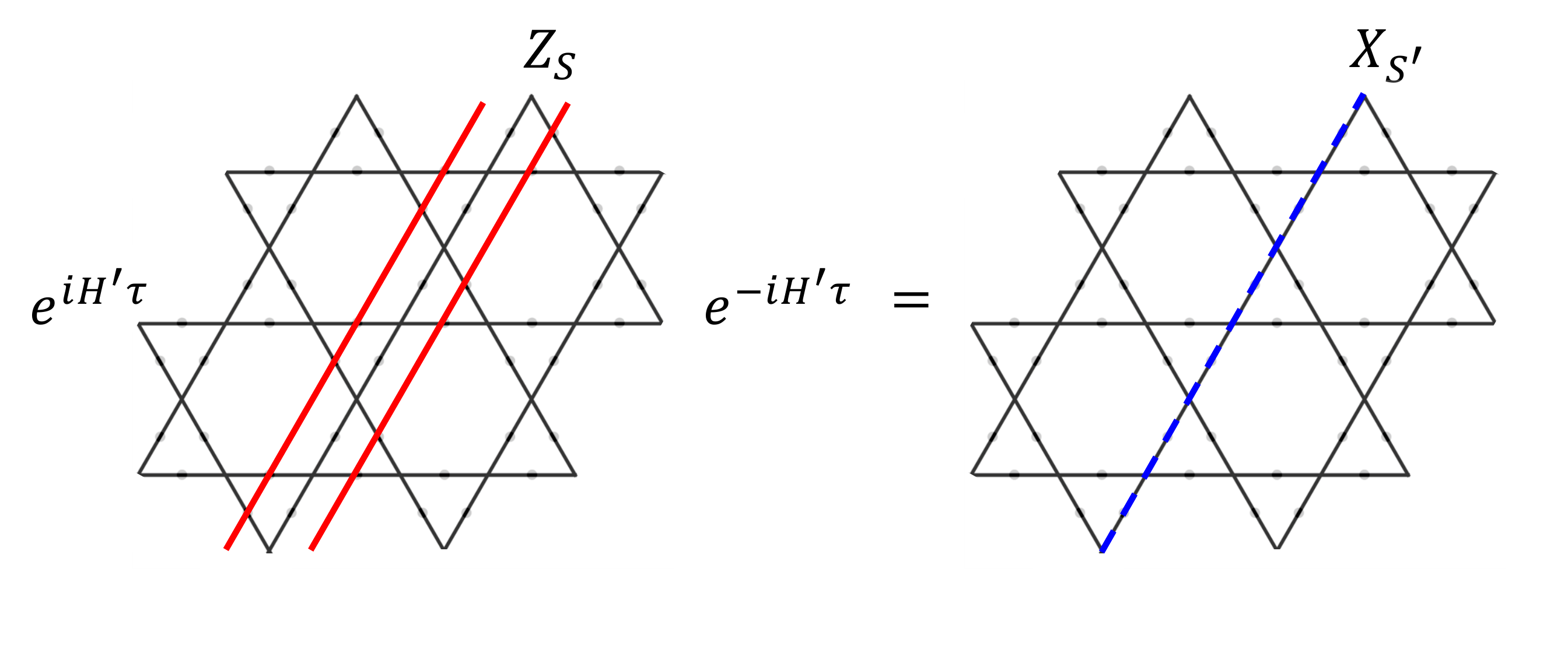} 
    \caption{The strings $S$ and $S'$ for the string operators $\bm{Z}_S$ and $\bm{X}_{S'}$, respectively, related by Eq.\ \eqref{eq:3a}.}
    \label{fig:6a}
\end{figure}
This duality follows from
\begin{equation} 
e^{i\tau H'} \bm{Z}_S e^{-i\tau H'}  =  \bm{X}_{S'} \ .
\label{eq:3a}
\end{equation} 
Figure \ref{fig:6a} shows the strings $S,S'$. To see this, concentrate on one of the triangles with vertices at the sites $i_1,i_2,i_3$ in the lattice that the string $S$ goes through. If it cuts the triangle at $i_1,i_2$, it contains a factor $Z_{i_1}Z_{i_2}$. Using \eqref{eq:1}, we obtain
\be e^{i\tau H_{\mathcal{T}}} Z_{i_1}Z_{i_2} e^{-i\tau H_{\mathcal{T}}} = X_{i_3} \ee 
By including the contributions of all triangles along the string, we deduce Eq.\ \eqref{eq:3a}.

For the time evolution we used the set of parameters  for the van der Waals interaction with $R_b  = 1.53 a $ and $R_{\text{trunc}}  = a $, following Ref.\ \cite{bib:12}.  Thus, the time evolution operator acted only within the triangles of the Ruby lattice. 
 
\begin{figure}[ht!]
    \centering
\includegraphics[width=0.475\textwidth]{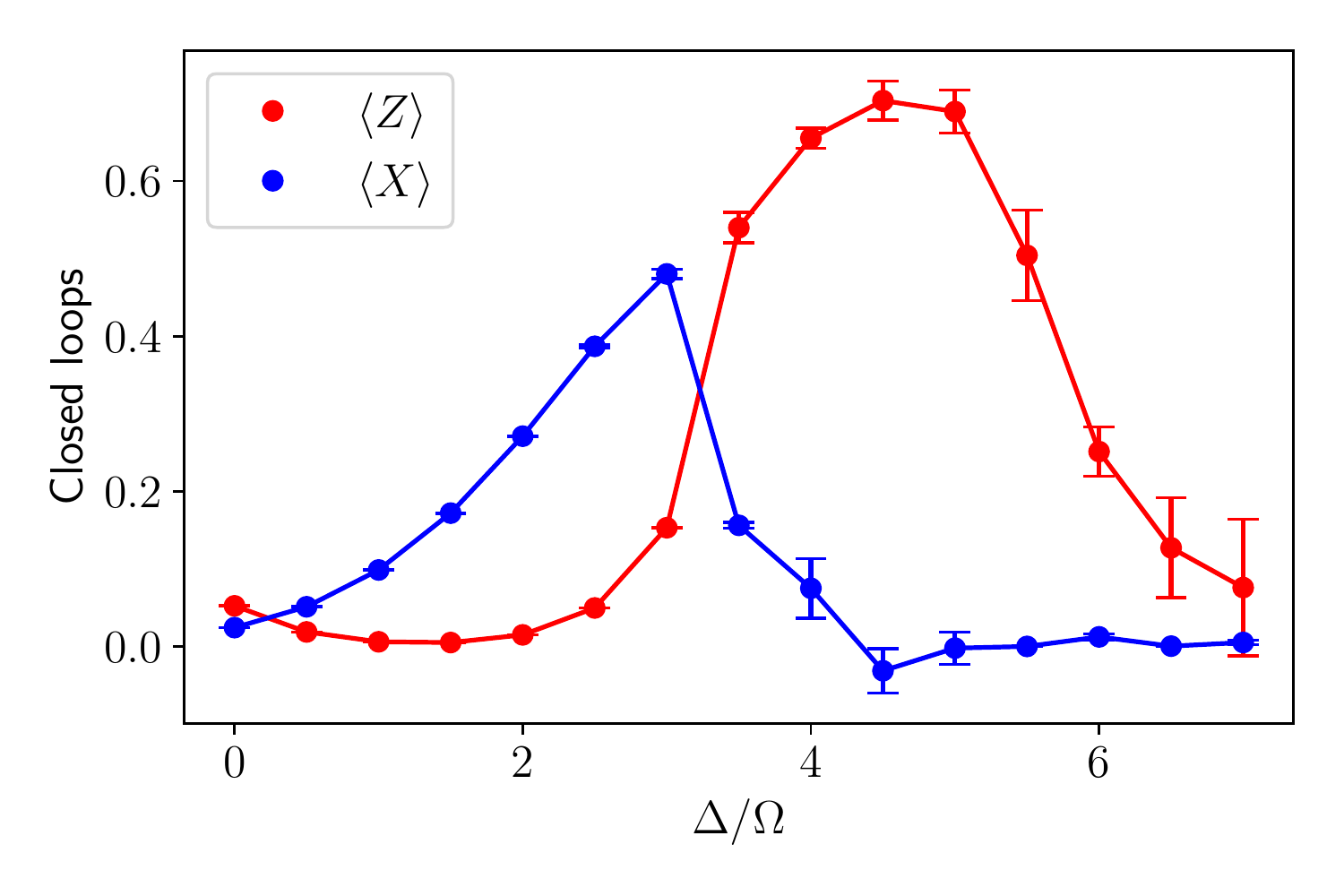} 
    \caption{Expectation values of closed $\bm{X}$ and $\bm{Z}$ strings as a function of $\Delta / \Omega $, obtained using iDMRG averaged over all single-hexagon loops in a $6\times 4 $ infinite cylinder system.  } 
    \label{fig:III1}
\end{figure}

To confirm that the system is in the spin liquid phase, we calculated the expectation values of open and closed $\bm{X}$  and $\bm{Z}$ strings for a range of $\Delta/\Omega$ values in a system without punctures. As discussed in \cite{bib:12}, the system supports 3 phases: trivial, spin liquid, and valence bond solid (VBS). The trivial phase appears when the ratio $\Delta/\Omega$ is small, and is characterized by finite open $\bm{X}$ strings and vanishing open $\bm{Z}$ strings, with the closed $\bm{X}$ and $\bm{Z}$ strings vanishing. For large $\Delta/\Omega $ values, the VBS state is realized, which has finite open $\bm{Z}$ strings and vanishing open $\bm{X}$ strings, with the closed $\bm{X}$ and $\bm{Z}$ strings also vanishing. For intermediate values of $\Delta/\Omega$, we obtain the spin liquid phase with vanishing open $\bm{X}$ and $\bm{Z}$ strings and finite closed $\bm{X}$ and $\bm{Z} $ strings. These results are plotted in Figure \ref{fig:III1}. 

%\begin{figure}[ht!]
%    \centering
%\includegraphics[width=0.50\textwidth]{0P_4x4_iXC_Chi100_Z_X_hex.pdf} 
%    \caption{Closed hexagonal string operators.} 
%    \label{fig:III1}
%\end{figure} 
% 

For numerical results, we restrict attention to a single point in the space of parameters, $\Delta/\Omega=3.5$. This is a global choice for all sites except at the boundaries of punctures. We set the detuning at the sites on the $e$-condensed boundaries to $\Delta' = 0.48 \Delta $. We confirm that the system supports mixed-boundary punctures by measuring the expectation values of $\bm{Z}$ and $\bm{X } $ strings that start and end on the two boundary types, as shown in Figure \ref{fig:IIlast}. Numerical results are given in Table \ref{tab:2a}. 

     \begin{table} 
    \centering 
    \begin{ruledtabular}
\begin{tabular}{lcccc}   % $\ket{\mathbb{I } } $ & 
 Operator & Expectation Value \\ 
 \colrule 
   $ \bm{Z}_1  $   & $0.0259 \pm 0.0129$ \\ % 0.0302 \pm 0.0117  \\ % 3 0.142  
  $ \bm{X}_1  $     & $1.086 \pm 0.0612$ \\ % 3 0.142  
   $ \bm{Z}_2  $   & $0.974 \pm 0.196$  \\  
   $ \bm{X}_2  $    & $0.300 \pm  0.0496$  \\  
\end{tabular} 
\end{ruledtabular}
\caption{ Numerical results for normalized expectation values for the string operators shown in Figure \ref{fig:IIlast} using iDMRG.   } \label{tab:2a} 
    \end{table}  
 
We now realize the ground states described in the previous section.  In particular, we show results for $\ket{e}_{12}$ and $\ket{m}_{12}$ ground states on the 2-puncture system with mixed boundaries, which are relevant for the non-Abelian braiding statistics. To differentiate between the various ground states involving punctures $p_1,p_2$ ($\ket{e }_{12 }  $, $\ket{m }_{12 }  $, $\ket{\pm }_{12 }  $),  we measure the set of non-local string operators $\{ \bm{Z}_C,\bm{X}_{C'},\bm{Z}_{S}, \bm{X}_{S'} \} $. Again, we take advantage of the periodic boundary conditions and measure the operators shown in Figure \ref{fig:II5}.  These results are shown in Table \ref{tab:2}. Note that the values for $\langle \bm{Z}_{S} \rangle $ should be 0 but have high standard error values, such that they are consistent with 0 with 2 standard errors. 
 
We normalize the expectation values of our non-local operators by calculating the joint expectation value of two strings.  For example, for $\bm{Z}_S$, we measure $\langle \bm{Z}_{S_i}\rangle$ along multiple strings $S_i$, and we also calculate $\langle \bm{Z}_{S_i}\bm{Z}_{S_{j}} \rangle $ for neighboring strings $S_i $ and $S_{j} $. The normalized expectation value of the string operator $\bm{Z}_S$ is obtained by averaging  $\langle \bm{Z}_{S_i} \rangle/\sqrt{\langle \bm{Z}_{S_i}\bm{Z}_{S_j }\rangle } $ over strings $i$ and $j$.  %\textcolor{red}{logical [what's logical about them?]} operators $\bm{Z}_{C }  $, $\bm{X}_{C } $ , $\bm{Z}_{S } $ and $\bm{X}_{S }  $ (see Fig.\ \textcolor{red}{[What is the right figure?]}). 
%The operator $\bm{Z}_{C }  $ ($ \bm{X}_{C }  $) is defined along a closed loop $C$ around one of the punctures, measuring whether an $e $ ($m $) anyon is enclosed. The operator $\bm{Z}_S $ ($ \bm{X}_S $) is defined along a string $S$ connecting the $m $ ($e $) boundaries of the two punctures. It is an open string containing $m $ ($e $) anyons at the two ends. Its expectation value is the overlap of the ground state with two $m $ ($e $) anyons added to the punctures with the initial ground state. \textcolor{red}{[Not clear. What is the initial ground state?]} % We work similarly for the states involving punctures $p_3,p_4$.  

\begin{table}[b!]
\begin{ruledtabular}
\begin{tabular}{lcccc} &  $\ket{e}$ & $\ket{m}$ \\ % $\ket{\mathbb{I } } $ & 
\colrule 
$\bm{Z}_{C_1}$ &  $1.007  \pm 0.0487$  & $-1.154  \pm 0.0990$  \\ % -0.938 
$\bm{Z}_{C_2}$ & $-1.0266 \pm 0.0523$  & $-0.931  \pm 0.000849$ \\ % 1.066 
$\bm{Z}_{C_3}$ & $1.015 \pm 0.0543$  & $-0.999  \pm 0.0633$ \\ % -0.963 
$\bm{X}_{C'_1} $ & $0.826 \pm 0.136$  & $1.261 \pm 0.0313$  \\ % 1.047
$\bm{X}_{C'_2} $ & $1.111 \pm 0.0994$  & $-0.811 \pm 0.312$  \\ % 0.687
$\bm{X}_{C'_3} $  & $1.072 \pm 0.0224$  & $1.097 \pm 0.0535$ \\ % 0.891 
$\bm{Z}_{S}$ & $-0.309 \pm 0.301$  & $-0.495  \pm 0.329$  \\ % rerunning 
$\bm{X}_{S'}$ & $-0.000183 \pm 0.000113$ & $-0.0132  \pm 0.00648$ \\ 
\end{tabular} 
\end{ruledtabular}
\caption{\label{tab:2} Normalized expectation values for the operators shown in Figure \ref{fig:II5} for ground states   $\ket{e}$ and $\ket{m}$ obtained using iDMRG. }
\end{table} 
%First, we considered a lattice with $L_x=4$ and $L_y=4$ on an infinite cylinder. 

% The regions of sites on the lattice with detuning $\Delta $ are in the spin liquid phase, where each triangle on the lattice has either 0 or 1 excitation, that is, $\langle n \rangle = 0 , 1 $. However, when sites on the $X$ boundaries have detuning $\Delta ' $, the sites can have $\langle n \rangle =0, 1/2 $. Then, $\langle \sigma^{z}  \rangle = 2 \langle n \rangle  - 1 $ vanishes, and the $\langle Z \rangle $ for loops starting and ending on these $X $ boundaries also vanish. This behavior can be seen in the Rydberg excitation plot in FIG?.  
 
\section{Quantum computation}\label{sec:IV}

To realize the fusion rules of Ising anyons, $e \times m=f $  with $f $ a fermion, with the $e $ and $m $ anyons fusing with themselves to form a vacuum state $\mathbb{I}$ ($e\times e = m\times m = \mathbb{I}$), we consider a system with four mixed-boundary punctures, $p_1,p_2,p_3,p_4$.
%The punctured Ruby lattice of Rydberg atoms is shown in Figure  \textcolor{red}{[What is the right figure?]}. 
Two $e$-anyons ($m$-anyons) can be created between the punctures $p_1$ and $p_2$ by connecting their rough (smooth) boundaries with a $\bm{Z}_S$ ($\bm{X}_S$) string, and similarly for the $p_3$ and $p_4$ punctures. This allows us to define the states $\ket{\pm}_{12}$ (Eq.\ \eqref{eq:2a}) and, similarly, $\ket{\pm}_{34}$, respectively. 
 
We restrict attention to the two-dimensional Hilbert space spanned by $\{ \ket{+}_{12} \ket{+}_{34} , \ket{-}_{12} \ket{-}_{34} \}$. These basis states can be written in terms of four-puncture states corresponding to $\mathbb{I}$ and $f$, respectively,
\begin{equation} 
\begin{split} 
\ket{\mathbb{I} }_{12 34 }  = \frac{1}{\sqrt{2}} \left( \ket{e }_{12 } \ket{e }_{34 }  + \ket{m }_{12 } \ket{m }_{34 } \right)  \\ 
\ket{f }_{12 34 } = \frac{1}{\sqrt{2}} \left( \ket{e }_{12 } \ket{m }_{34 }  + \ket{m }_{12 } \ket{e }_{34 }  \right)
\end{split} 
\label{eq:4a} 
\end{equation} 
as 
\begin{equation} 
\ket{\pm }_{12}  \ket{\pm }_{34}  = \frac{1 }{\sqrt{2} } \left( \ket{\mathbb{I} }_{12 34 }  \pm \ket{f }_{12 34 } \right)  
\label{eq:5} 
\end{equation} 
It follows that the transformation between the basis states $\ket{\pm}_{12} \ket{\pm}_{34}$ and $\{ \ket{\mathbb{I}}_{1234} , \ket{f}_{1234} \}$ is given by the matrix
\begin{equation} 
F = \frac{1 }{\sqrt{2 } } \begin{pmatrix} 1 & 1 \\ 1 & -1  \end{pmatrix}  
\label{eq:6} 
\end{equation} 
which is the fusion matrix in the Ising model.

%Now, we consider the braiding statistics of this system with four mixed-boundary punctures. we can braid $p_1 $ around $p_3 $. We have the $R$-matrix elements $R_{em}=R_{me}=-1 $ and $R_{ee}=R_{mm}=1 $, which give the phase factors picked up by exchanging the Abelian $e$ and $m$ anyons. \textcolor{red}{[There is no mention of the $R$-matrix before. A short discussion is in order.]}  We obtain $\ket{\mathbb{I}} \to \ket{\mathbb{I}}$ and $\ket{f}\to - \ket{f}$. Thus, the braiding operation acts as a Pauli $Z$ matrix in the basis $\{ \ket{\mathbb{I}} , \ket{f} \}$.

For a pair of puncture, having an odd number of $\bm{X}$ ($\bm{Z}$) strings from one $e$ ($m$) condensed boundary to another creates an $e$ ($m$) state. It follows that we can create all states of interest by creating $\bm{X}$ and $\bm{Z}$ strings between boundaries as needed. To this end, we  introduce an additional array of ancilla atoms, following the scheme proposed in \cite{bib:25}. The ancilla array is constructed using atoms that are of a different type from those in the Ruby lattice (e.g., ${}^{23}$Na). 
We can then implement the two-qubit unitary (controlled-$Z$) gate between an ancilla atom $a$ (control) and and a Rydberg atom at site $i$ in the Ruby lattice (target),
\begin{equation} 
    CZ_i^a =\ket{0}_a\bra{0}\otimes\mathbb{I}+\ket{1}_a\bra{1}\otimes Z_i    
\end{equation} 
The $CZ_i^a$ gate can be implemented through a number of ways\cite{bib:27}\cite{bib:28}, but we are primarily proposing the use of cold collisions between the ancilla and Rydberg atoms\cite{bib:26}. This is done by bringing the ancilla atoms close to the code atoms so that their wave packets overlap for a brief time. This causes the wavefunctions of the atoms to acquire a relative phase 
whose amount can be controlled by the interaction time. The relationship between the interaction time and the gate fidelity has also been investigated; see \cite{bib:26} for details.
 
We can use the aforementioned gate to create the $\bm{Z}$ string. To see this, consider the action of the $\bm{Z}$ string on the triangles of the Kagome lattice as described in \cite{bib:12}. Whenever a $\bm{Z}$ string passes through a dimer, it acquires a phase of $-1$. The $CZ$ gate reproduces this effect if the ancilla atom is projected onto the $\ket{1}_a$ state. We can then use the ancilla to apply a series of $Z_i$ gates along a string $S$ ($i\in S$) consisting of segments perpendicular to the links of the Kagome lattice and connecting one $m$-boundary on one puncture to the same boundary on a different puncture. This gives a $\bm{Z}_S$ string. We obtain the controlled-$\bm{Z}_S$ gate
\be C\bm{Z}_S^a = \prod_{i\in S} CZ_i^a = \ket{0}_a\bra{0}\otimes\mathbb{I}+\ket{1}_a\bra{1}\otimes \bm{Z}_S  \label{eq:CZS}\ee
For the $\bm{X}$-strings, we make use of Eq (\ref{eq:3a}). One simply has to identify the appropriate conjugate paths for the required $\bm{X}$ string as shown in fig \ref{fig:6a} and time evolve appropriately.

\begin{equation}
    C\bm{X}_{S'}^a = e^{i\tau H'}\cdot C\bm{Z}_S^a\cdot  e^{-i\tau H'}
    \label{eq:CXS}
\end{equation} 

We can create the superposition state in Eq (\ref{eq:2a}) by using an ancilla pair in the $\frac{1}{\sqrt{2}} (\ket{01} \pm \ket{10} )$ bell state. Each ancilla creates a $\bm{Z}_S$ along their defined paths. The first ancilla creates the $C\bm{Z}_{S_1}^{a_1}$ string between the $m$-condensed boundaries, then according to Eq (\ref{eq:CXS}) we time evolve the system and apply the requisite $C\bm{Z}_{S_2}^{a_2}$ along the appropriate path(s) connecting the $e$-boundaries and time evolve again. This creates a superposition of the $\bm{X}$ and $\bm{Z}$ strings, giving us the superposition states we want.

%\begin{figure}[ht!]
%    \centering 
%\includegraphics[width=0.45\textwidth]{AncillaQubits.pdf} 
%    \caption{.} 
%    \label{fig:IVAQ} 
%\end{figure} 

Now, consider a Ruby lattice with two mixed-boundary punctures $p_1,p_2$ in the topologically trivial ground state $\ket{\mathbb{I}}$. To create a state in the two-dimensional Hilbert space spanned by $\{ \ket{+}_{12}  , \ket{-}_{12} \}$, we need 2 ancilla qubits to use as controls to apply the string operators $\bm{X}_{S_1}, \bm{Z}_{S_2}$, where $S_1, S_2$ are paths joining the two punctures. We obtain the state
\be \ket{\Phi_2} = C\bm{X}_{S_1}^{a_1}\cdot C\bm{Z}_{S_2}^{a_2}  \ket{\psi_2}_{a_1a_2} \ket{\mathbb{I}}_{12} \ee
where $\ket{\psi_2}$ is the initial state of the two ancilla qubits that can be chosen at will. The state $\ket{\Phi_2}$ is a superposition of several states including unwanted states. To ensure that only the states $ \ket{\pm}_{12} $ contribute, we perform measurements on the ancilla qubits that project onto a state in the span of $\{ \ket{01}_{a_1a_2} , \ket{10}_{a_1a_2} \}$. Then the state $\ket{\Phi_2}$ can be written as the superposition
\be \ket{\Phi_2} = \sum_{\sigma = \pm} c_\sigma \ket{\Psi^\sigma}_{a_1a_2} \ket{\sigma}_{12} \ee
where
\be\label{eq:Bs} \ket{\Psi^\pm} = \frac{1}{\sqrt{2}} (\ket{01} \pm \ket{10} ) \ee
are Bell states, and the coefficients $c_\pm$ depend on the choice of the initial state $\ket{\psi_2}$ of the ancilla qubits. Their exact value is not important, allowing us freedom in preparing the initial ancilla state. A measurement of the ancilla qubits in the Bell state basis initializes the system in one of the states $\ket{\pm}_{12}$ (Eq.\ \eqref{eq:2a}). %\textcolor{red}{[Explain how this measurement is done. One way is by applying $H_{a_1}\cdot \textrm{CNOT}_{a_1a_2}$ which maps $\ket{\Psi^+} \mapsto \ket{01}$ and $\ket{\Psi^-}\mapsto \ket{11}$, and then measuring the $a_1$ ancilla qubit. Can these gates be implemented? If so, how?]}

Next, we consider a Ruby lattice with four mixed-boundary punctures $p_1,p_2,p_3,p_4$ in the topologically trivial ground state $\ket{\mathbb{I}}$. To create a state in the two-dimensional Hilbert space spanned by $\{ \ket{+}_{12} \ket{+}_{34} , \ket{-}_{12} \ket{-}_{34} \}$, we need 4 ancilla qubits to use as controls to apply the string operators $\bm{X}_{S_1}, \bm{Z}_{S_2}, \bm{X}_{S_3}, \bm{Z}_{S_4}$, where $S_1, S_2$ ($S_3,S_4$) are paths joining punctures $p_1,p_2$ ($p_3,p_4$). We obtain the state
\be \ket{\Phi_4} = C\bm{X}_{S_1}^{a_1}\cdot C\bm{Z}_{S_2}^{a_2} \cdot C\bm{X}_{S_3}^{a_3} \cdot C\bm{Z}_{S_4}^{a_4} \ket{\psi_4}_{a_1a_2a_3a_4} \ket{\mathbb{I}}_{1234} \ee
where $\ket{\psi_4}$ is the initial state of the four ancilla qubits whose specific form is not important. To ensure that only the states $ \ket{\pm}_{12} \ket{\pm}_{34} $ contribute, we perform measurements on the ancilla qubits that project onto a state in the span of $\{ \ket{\Psi^+}_{a_1a_2} \ket{\Psi^+}_{a_3a_4} , \ket{\Psi^-}_{a_1a_2} \ket{\Psi^-}_{a_3a_4} \}$, This is equivalent to demanding that the projection of $\ket{\Phi_4}$ be stabilized by the commuting operators $X_{a_1} X_{a_2} X_{a_3} X_{a_4}$, $Z_{a_1} Z_{a_2}$, and $Z_{a_3} Z_{a_4}$, specifically,
\be\label{eq:16} \begin{split} X_{a_1} X_{a_2} X_{a_3} X_{a_4} \ket{\psi_4} &= \ket{\psi_4}  \\ Z_{a_1} Z_{a_2} \ket{\psi_4}  &= -\ket{\psi_4} \\  Z_{a_3} Z_{a_4} \ket{\psi_4} &= -\ket{\psi_4} \end{split}\ee
To project onto one of the states $ \ket{\pm}_{12} \ket{\pm}_{34} $, we need to additionally measure $X_{a_1} X_{a_2}$ (or, equivalently, $X_{a_3}X_{a_4}$), because they belong to different eigenvalues, since $X_{a_1} X_{a_2} \ket{\Psi^\pm}_{a_1a_2} = \pm \ket{\Psi^\pm}_{a_1a_2}$. If, instead, we are interested in one of the states $\{ \ket{\mathbb{I}}_{1234}, \ket{f}_{1234} \}$, we need to measure $Z_{a_2} Z_{a_3}$ (or $Z_{a_1}Z_{a_4}$) whose eigenstates are $\frac{1}{\sqrt{2}} (\ket{\Psi^+}_{a_1a_2}\ket{\Psi^+}_{a_3a_4} \pm \ket{\Psi^-}_{a_1a_2}\ket{\Psi^-}_{a_3a_4})$. 
%Notice that these two operators are related by
%\be Z_{a_2} Z_{a_3} = \mathcal{U}\cdot X_{a_1}  X_{a_2} \cdot \mathcal{U} \ , \ee 
%where
%\be \mathcal{U} = \mathrm{SWAP}_{a_1a_3} H_{a_1}H_{a_2}H_{a_3}H_{a_4}\ee
%Notice that $\mathcal{U}^\dagger = \mathcal{U}$, $\mathcal{U}^2 = \mathbb{I}$.
%Thus, we can switch between the two bases by applying $\mathcal{U}$ which amounts to swapping (braiding) ancilla qubits 1 and 3. 

In a quantum computation we need to prepare an initial state, apply gates, and then perform measurements on the final state. To prepare the initial state, we may perform measurements on the ancilla qubits bringing the state of the system in the form
\be \ket{\Phi_4} = \sum_{\sigma = \pm} c_\sigma \ket{\Psi^\sigma}_{a_1a_2} \ket{\Psi^\sigma}_{a_3a_4} \ket{\sigma}_{12} \ket{\sigma}_{34} \ee
where the coefficients $c_\pm$ depend on the choice of the initial state $\ket{\psi_4}$ of the ancilla qubits. Their exact value is not important. While an appropriate measurement of the ancilla qubits will yield the desired initial state, it decouples the ancilla qubits and can no longer be used in the quantum computation. To remedy this, 
%we attach an additional ancilla qubit labeled $b$ and entangle it with the other ancilla qubits to obtain the state
%\be \sum_{\sigma = \pm} c_\sigma \ket{\Psi^\sigma}_{a_1a_2} \ket{\Psi^\sigma}_{a_3a_4} \ket{\sigma}_b\ket{\sigma}_{12} \ket{\sigma}_{34} \ee
instead of fixing the initial state by an ancilla measurement, we first apply the quantum gates and then measure all ancilla qubits. %The measurement on $b$ is appropriate for the desired initial state, whereas measurement of the remaining ancilla qubits yields the final state after the quantum computation.
Indeed, let $\ket{\sigma}_{12} \ket{\sigma}_{34}$ be the desired initial state of the system for a given $\sigma$, and $U$ the unitary that implements the quantum computation. The final state is
\be\label{eq:fin} \sum_{\sigma' = \pm} U^{\sigma\sigma'} \ket{\sigma'}_{12} \ket{\sigma'}_{34} \ee
where $U^{\sigma\sigma'}$ is a $2\times2$ matrix. To implement $U$, we construct a dual unitary $\widetilde{U}$ acting on the ancilla qubits such that
\be \widetilde{U} \ket{\Psi^\sigma}_{a_1a_2} \ket{\Psi^\sigma}_{a_3a_4} = \sum_{\sigma' = \pm} \widetilde{U}^{\sigma\sigma'} \ket{\Psi^{\sigma'}}_{a_1a_2} \ket{\Psi^{\sigma'}}_{a_3a_4}  \ee
where the $2\times2$ matrix $\widetilde{U}^{\sigma\sigma'}$ is the transpose of $U^{\sigma\sigma'}$. By acting with $\widetilde{U}$ on the state $\ket{\Phi_4}$ and then measuring the ancilla qubits with outcome $\sigma$, the state of the system collapses to the desired final state \eqref{eq:fin}.

We can realize the braiding group of Ising anyons by acting with the dual exchange matrices
\be\label{eq:17} \widetilde{R}_{12} = e^{i\frac{\pi}{4} X_{a_1} X_{a_2}} \ , \ \widetilde{R}_{23} = e^{-i\frac{\pi}{4} Z_{a_2} Z_{a_3}} \ , \ee
as well as $\widetilde{R}_{34} = e^{i\frac{\pi}{4} X_{a_3} X_{a_4}} \ , \ \widetilde{R}_{41} = e^{-i\frac{\pi}{4} Z_{a_4} Z_{a_1}}$, on the ancilla qubits. The latter reduce to those in \eqref{eq:17} in the subspace defined by \eqref{eq:16}. After measuring the ancilla qubits, their action amounts to exchange matrices $R_{ij}$ on the system of anyons on the Ruby lattice matching those of Ising anyons.

The Ising fusion matrix $F$ (Eq.\ \eqref{eq:6}) is similarly implemented via its dual
\be \widetilde{F} = \widetilde{R}_{12}^{-1} \widetilde{R}_{23} \widetilde{R}_{12}^{-1} \ee
acting on the ancilla qubits.
Also, the logical $X$ is implemented via $\widetilde{X} \equiv (\widetilde{R}_{23})^2$, and can also be obtained by braiding punctures $p_1$ and $p_3$ \cite{bib:21}.

To extend the above results to two logical qubits, we consider a system with six mixed-boundary punctures, $p_1,\dots,p_6$ that requires six ancilla qubits $a_1,\dots, a_6$ for control. The four-dimensional Hilbert space of the two logical qubits is obtained by restricting the state of the ancilla qubits $\ket{\psi_6}$ with the stabilizer constraints
\be\label{eq:16a} \begin{split} X_{a_1} X_{a_2} X_{a_3} X_{a_4} X_{a_5} X_{a_6} \ket{\psi_6} &= \ket{\psi_6}  \\ Z_{a_1} Z_{a_2} \ket{\psi_6}  &= -\ket{\psi_6} \\  Z_{a_3} Z_{a_4} \ket{\psi_6} &= -\ket{\psi_6} \\  Z_{a_5} Z_{a_6} \ket{\psi_6} &= -\ket{\psi_6} \end{split}\ee
It is spanned by the basis \be\{ \ket{\psi_6^{+++}} , \ket{\psi_6^{+--}} ,\ket{\psi_6^{-+-}} ,\ket{\psi_6^{--+}} \} \ , \ee where  $\ket{\psi_6^{\sigma_1\sigma_2\sigma_3}} =\ket{\Psi^{\sigma_1}}_{a_1a_2}\ket{\Psi^{\sigma_2}}_{a_3a_4}\ket{\Psi^{\sigma_3}}_{a_5a_6} $ ($\sigma_{1,2,3} = \pm$).

The Ising anyon braiding group is realized by defining dual exchange matrices $\widetilde{R}_{ij}^{(6)}$ acting on the ancilla qubits, as before. The exchange matrices on the Ruby lattice anyons, ${R}_{12}^{(6)}$ and ${R}_{23}^{(6)}$ remain single-qubit matrices. Indeed, we easily obtain the decomposition of the $4\times4$ matrices on the lattice anyons, ${R}_{12}^{(6)} = {R}_{12}\otimes \mathbb{I}$ and ${R}_{23}^{(6)} = {R}_{23}\otimes \mathbb{I}$ in terms of the $2\times2$ matrices \eqref{eq:17}. This is not the case for the exchange matrix ${R}_{56}^{(6)}$ which is a non-trivial two-qubit matrix. We obtain
\be R_{34}^{(6)} = e^{i\frac{\pi}{4} X_{a_3} X_{a_4}} = \ket{0}_L \bra{0} \otimes R_{12} + i \ket{1}_L \bra{1} \otimes R_{12}^{-1} \ee
implementing a controlled-Z gate (since $R_{12}^2 = Z$ on a logical qubit).

Extension to a larger number of logical qubits is straightforward. For $N$ logical qubits, we consider a system of $2N+2$ mixed-boundary punctures $p_1,\dots,p_{2N+2}$, and an equal number of ancilla qubits, $a_1,\dots, a_{2N+2}$. We restrict the Hilbert space to $2^N$ dimensions by imposing the stabilizer conditions
\be\label{eq:25} \begin{split}
    \prod_{i=1}^{2N+2} X_{a_i} \ket{\psi_{2N+2}} &= \ket{\psi_{2N+2}} \\
    Z_{a_{2i-1}}Z_{a_{2i}} \ket{\psi_{2N+2}} &= -\ket{\psi_{2N+2}} \ \ (i=1,\dots,N+1)
\end{split} \ee
on the ancilla state $\ket{\psi_{2N+2}}$. We define quantum gates by using the dual exchange matrices on the ancilla qubits
\be \widetilde{R}_{2i-1\, 2i}^{(2N+2)} = e^{i\frac{\pi}{4} X_{a_{2i-1}} X_{a_{2i}}} \ , \ \ \widetilde{R}_{2i\, 2i+1}^{(2N+2)} = e^{-i\frac{\pi}{4} Z_{a_{2i}} Z_{a_{2i+1}}} \ee
which commute with the stabilizers \eqref{eq:25}.

\section{Conclusion}\label{sec:V} 
We have devised a scheme to obtain non-Abelian braiding statistics using mixed boundary punctures in a spin liquid of Rydberg atoms on a Ruby lattice. The spin liquid is diagnosed by computing closed (open) string operators which we find to be finite (vanishing). We confirm that the punctures produced in the system realize both the $e$ and $m$ boundary types by observing non-vanishing $\bm{X}_S$ and $\bm{Z}_{S'}$ expectation values along open strings $S,S'$ ending on the boundaries of punctures.  The system of the spin liquid with two mixed-boundary punctures can support a set of four (quasi)-degenerate ground states, $\{ \ket{\mathbb{I}}, \ket{e}, \ket{m}, \ket{\epsilon}\}$ identified by measuring the non-local loop operators $\bm{Z}_C,\bm{X}_{C'},\bm{Z}_{S},\bm{X}_{S'}$, where $S,S'$ join two punctures and $C,C'$ are loops around a puncture. 
 
We showed that superposition states, such as $\ket{\pm} = 1/\sqrt{2}(\ket{e} \pm \ket{m} ) $ (Eq.\ \eqref{eq:2a}),  can be used to define states that realize non-Abelian Ising braiding statistics in a multi-puncture system. %implementing the $X$ gate on our logical basis. 
We discussed an experimental scheme using cold collisions involving ancilla atoms to prepare topologically nontrivial ground states from the trivial state $\ket{\mathbb{I}} $. We obtained $N$ logical qubits in a Ruby lattice with $2N + 2$
mixed-boundary punctures, and an equal number of ancilla atoms that we added for control. The $N$-qubit
Hilbert space was obtained by constraining the state of the
ancilla atoms using a set of stabilizing conditions. Quantum gates could then be implemented using a set of gates acting
on the ancilla atoms that commuted with the stabilizers and formed a realization of the braiding group of
non-Abelian Ising anyons.  

It would be interesting to explore the possibility of implementing Ising braiding statistics without the addition of ancilla atoms, e.g., by engineering quantum gates acting directly on the Ruby lattice without the need of control operations by the ancilla atoms. Extension of our results to realize different non-Abelian braiding, such as Fibonacci anyons, would also be of interest. Work in this direction is in progress.
 
\acknowledgements
Research funded by the National Science Foundation under award DGE-2152168 and  the Army Research Office under award W911NF-19-1-0397.

\end{document}